\documentclass[aps,showpacs,nofootinbib]{revtex4}
\usepackage{amsmath} 
\usepackage{amssymb}
\usepackage{graphicx} 
\usepackage{color} 
\usepackage{epstopdf}

\begin{document}

\newcommand{\blue}[1]{\textcolor{blue}{#1}}
\newcommand{\new}{\blue}
\newcommand{\green}[1]{\textcolor{green}{#1}}
\newcommand{\modif}{\green}
\newcommand{\red}[1]{\textcolor{red}{#1}}
\newcommand{\attention}{\red}


\title{Global characteristics of the medium produced in ultra-high energy cosmic ray collisions}

\author{V. A. Okorokov} \email{VAOkorokov@mephi.ru; Okorokov@bnl.gov}
\affiliation{National Research Nuclear University MEPhI (Moscow
Engineering Physics Institute), Kashirskoe highway 31, 115409
Moscow, Russia}

\date{\today}

\begin{abstract}
Estimations of some geometrical and bulk parameters are
presented for the matter produced in various type collisions with
ultra-high energy cosmic ray (UHECR) particles. Results for
multiplicity density at midrapidity, decoupling time, and energy
density are discussed for small and larger collision systems. Based
on the analytic functions suggested previously elsewhere,
estimations for a wide set of space-time quantities are obtained
for emission region created in various particle collisions at
energies of UHECR. The space particle densities at freeze-out are
derived also and allow the possibility of novel
features for secondary particle production like Bose--Einstein
condensation at least for nuclear interactions with UHECR
particles. The estimations obtained for global and geometrical parameters
indicate the creation of deconfined quark-gluon matter with
large enough volume and lifetime even in light nuclear collisions
at UHECR energies. These quantitative results can be important for both
the future collider experiments at center-of-mass energy frontier
and the improvement of the phenomenological models for development
of the cosmic ray cascades in ultra-high energy domain.
\end{abstract}

\pacs{98.70.Sa,
25.75.Nq
}

\maketitle

\section{Introduction}\label{sec:1}

The present projects of the future research facilities prove that
the accelerator physics in XXI century will be the physics in
$\mathcal{O}(100~\mbox{TeV})$ domain of the center-of-mass
energies. Measurements of interactions of cosmic ray particles
with ultra-high initial laboratory energies larger than  0.1--1
EeV with nuclei in the atmosphere allow the new unique
possibilities for study of multiparticle production processes at
energies (well) above not only the Large Hadron Collider -- LHC
range but foreseeable-future collider on Earth as well. Collisions
at such ultra-high energies can lead to creation of a strongly
interacting matter under extreme conditions. Due to the air
composition and main components of the ultra-high energy cosmic
rays (UHECR) the passage of UHECR particles through atmosphere can
be considered as collision of mostly small systems. Among the most
challenging problems for collider experiments is the study of the
quark-gluon matter created in such collisions. On the other hand,
the investigation of main properties of the final-state matter for
UHECR particle collisions with atmosphere can be useful for better
understanding of the origin and features of UHECR itself.
Therefore the estimations of global characteristics of the matter
produced in ultra-high energy cosmic ray collisions seems
important for both the experiments at present and, possibly, even
more for future colliders and the physics of cosmic rays.

\section{Observables and approximating functions}\label{sec:2}

Although the potential sources of UHECR are still quite far from
understanding, it is reasonable to suggest the same acceleration
mechanism for protons and heavier components of the UHECR which
affects the charged component of the nuclei. Some
electromagnetic fields can be such a mechanism but not the shock
waves from explosion processes. Therefore within the nucleus
$(A_{1}, Z_{1})$ incident on a particle at rest
$(A_{2}, Z_{2})$ with nucleon numbers
$A_{1}$, $A_{2}$ and charges $Z_{1}e$,
$Z_{2}e$ in conditions with electromagnetic field set for
protons of laboratory momentum $p_{p}$, the collision will be
characterized by the momentum per nucleon of the incoming nucleus
in laboratory reference system and the center-of-mass energy per
nucleon-nucleon pair
\begin{equation}
p_{N}=(Z_{1}/A_{1})p_{p}\,,~~~
\left.\sqrt{\smash[b]{s_{NN}}}\,\right|_{m_{N} \approx m_{p},\,
m_{p} \ll E_{p}} \approx
\sqrt{\smash[b]{(Z_{1}/A_{1})s_{pp}}}\,,
\label{eq:2.1}
\end{equation}
where $s_{NN/pp}=2m_{N/p}(E_{N/p}+m_{N/p})$ is standard Mandelstam
invariant variable, $E_{N/p}$ and $m_{N/p}$ is the energy in the
laboratory frame and mass of nucleon/proton
\cite{PDG-PRD-98-030001-2018}.

In the present study the global parameters are estimated with the help
of the extrapolation technique. The corresponding extrapolations use
the parameterizations that describe the existing experimental data
in dependence on $s$\footnote{Below the index ``NN / pp'' will be
omitted for this parameter for brevity if the statement is
applicable for both the $p+p$ and nucleus-nucleus collisions.}. It
should be noted that analytic functions given below were obtained
on the basis of the Standard Model (SM) without invoking any
hypothesis concerning contributions from physics beyond it. The
justification of the approach for ultra-high energy domain can be
found elsewhere \cite{Okorokov-PAN-82-134-2019}.

The energy dependence of the pseudorapidity ($\eta$) density of secondary charged particles produced at $\eta=0$ per nucleon-nucleon pair
can be approximated in particular by the universal power expression
\begin{equation}
\rho^{\eta} \equiv \xi^{-1}
\left.\bigl(dN_{\scriptsize{\mbox{ch}}}/d\eta\bigr)\right|_{\eta=0}
= a_{1}\varepsilon^{a_{2}},~~~ \varepsilon \equiv s/s_{0},
\label{eq:2.2}
\end{equation}
where $s_{0}=1$ GeV$^{2}$, $\xi \equiv 0.5\langle
N_{\scriptsize{\mbox{part}}} \rangle$, $\langle
N_{\scriptsize{\mbox{part}}}\rangle$ is the average number of
participants and $\langle N_{\scriptsize{\mbox{part}}}\rangle=2$
for $p+p$ collisions, free parameters are $a_{1}=0.75 \pm 0.06$,
$a_{2}=0.114 \pm 0.003$ for nonsingle diffractive $p+p$ collisions
at $\sqrt{\smash[b]{s_{pp}}} > 20$ GeV
\cite{Okorokov-arXiv-1606.08665-2016}, $a_{1}=1.94 \pm 0.08$,
$a_{2}=0.103 \pm 0.002$ for $p(d)+A$ interactions
\cite{EPJC-79-307-2019}, and $a_{1}=2.70 \pm 0.07$, $a_{2}=0.155
\pm 0.004$ for 0--5\% most central $A+A$ collisions
\cite{PRL-116-222302-2016,PLB-790-35-2019}. The total multiplicity of the charged
particles produced in nuclear collisions depends on $s$ and can be
approximated by the power function
\begin{equation}
\displaystyle \xi^{-1}
N_{\scriptsize{\mbox{ch}}}=a_{1}+a_{2}\varepsilon^{a_{3}},
\label{eq:2.3}
\end{equation}
where $a_{1}=-7.36 \pm 0.16$ $(-6.72 \pm 1.44)$, $a_{2}=6.97 \pm
0.12$ $(5.42 \pm 1.11)$, $a_{3}=0.133 \pm 0.001$ $(0.180 \pm
0.020)$ for $p+p$ (central $A+A$) collisions
\cite{PRD-93-054046-2016}.

Taking into account the experimental results for the central heavy ion
collisions \cite{PRC-71-034908-2005,PRL-109-152303-2012} one can
deduce the following approximation for the energy dependence of the
transverse energy ($E_{\,\scriptsize{T}}$) density at
$\eta=0$ normalized by participant pairs
\begin{equation}
\rho^{E}_{\scriptsize{T}} \equiv \xi^{-1}
\left.\bigl(dE_{\,\scriptsize{T}}/d\eta
\bigr)\right|_{\eta=0} = (0.46 \pm 0.16)\,\varepsilon^{0.200 \pm
0.005}. \label{eq:2.4}
\end{equation}
The Bjorken energy density and temperature of the matter created in the
collisions of UHECR particles in atmosphere depend on the time
duration ($\tau$) since the collision moment. The corresponding
estimations for most central collisions are following:
\begin{equation}
\rho^{E}_{\scriptsize{\mbox{Bj}}}(\tau) =
\bigl(S_{\perp}\tau\bigr)^{-1}\times
\left.\bigl(dE_{\,\scriptsize{T}}/d\eta
\bigr)\right|_{\eta=0} \approx \bigl(\pi
R^{2}\tau\bigr)^{-1}\times
\left.\bigl(dE_{\,\scriptsize{T}}/d\eta
\bigr)\right|_{\eta=0}, \label{eq:2.5}
\end{equation}
\begin{equation}
T(\tau)=\bigl[30\rho^{E}_{\scriptsize{\mbox{Bj}}}(\tau)/\pi^{2}n_{\scriptsize{\mbox{qg}}}\bigr]^{1/4},
\label{eq:2.6}
\end{equation}
where $S_{\perp}$ is the nuclei transverse overlap area, the
radius for incoming nucleus is estimated as the radius of
spherically-symmetric object $\forall\,A > 1: R = r_{0}A^{1/3}$
with $r_{0}=(1.25 \pm 0.05)$ fm \cite{book} and for $p+p$
collisions $R=(0.875 \pm 0.006)$ fm
\cite{Okorokov-IJMPA-33-1850077-2018}. Here the Stefan--Boltzmann
relation is used for derivation of the $T(\tau)$ and
$n_{\scriptsize{\mbox{qg}}}$ is the number of degrees of freedom of
the quark-gluon matter with $N_{f}=3$ active quark flavors within the present work.

The smooth energy dependences of the Bose--Einstein (BE) correlation
parameters allow the study of the geometry and space-time
extent of the emission region of secondary particles produced in
UHECR collisions. The linear scales (radii) of the homogeneity
region for the 3D Gaussian source of the charged pion pairs with
low relative momentum can be parameterized by the universal
function \cite{Okorokov-AHEP-2015-790646-2015}
\begin{equation}
f_{i}(\varepsilon)=a_{1}^{i}\bigl[1+a_{2}^{i}(\ln
\varepsilon)^{a_{3}^{i}}\bigr],~~~ i=s, o, l\label{eq:2.7}
\end{equation}
with the appropriate set of parameters for each direction $i$ of
the Pratt--Bertsch coordinate system. The volume of the
homogeneity region is calculated as
\cite{Okorokov-AHEP-2015-790646-2015}
\begin{equation}
V=(2\pi)^{3/2}R_{s}^{2}R_{l} \label{eq:2.8}
\end{equation}
and time duration since the collision moment until kinetic
freeze-out stage called also BE decoupling time and characterizing
the total duration of the longitudinal expansion of final-state
matter is \cite{PLB-696-328-2011}
\begin{equation}
\tau_{\scriptsize{\mbox{kin}}} \approx 0.875
\bigl[\left.\bigl(dN_{\scriptsize{\mbox{ch}}} /
d\eta\bigr)\right|_{\eta=0}\bigr]^{1/3}.\label{eq:2.9}
\end{equation}

\section{Results of extrapolations}\label{sect:3}

The energy range for protons in laboratory reference system
considered in the present paper is $E_{p}=10^{17}$--$10^{21}$ eV.
This range includes the energy domain corresponded to the
Greisen--Zatsepin--Kuzmin (GZK) limit
\cite{Greisen-PRL-16-748-1966} and somewhat expands it, taking
into account, on the one hand, both possible uncertainties of
theoretical estimations for the limit values for UHECR and
experimental results, namely, measurements of several events with
$E_{p} > 10^{20}$ eV and the absence of UHECR particle flux
attenuation up to $E_{p} \sim 10^{20.5}$ eV
\cite{Okorokov-PAN-81-508-2018} and, on the other hand, the
energies corresponding to the nominal value $\sqrt{\smash[b]{s_{pp}}}=14$ TeV
of the commissioned LHC as well as to the parameters for the main
international projects high energy LHC -- HE-LHC ($\sqrt{\smash[b]{s_{pp}}}=27$ TeV) and Future
Circular Collider -- FCC ($\sqrt{\smash[b]{s_{pp}}}=100$ TeV). Therefore the
estimations below can be useful for both the UHECR physics and the
collider experiments.

Here the following set of nuclei $\mathcal{G}_{Y} \equiv
\bigl\{\mathcal{G}_{Y}^{i}\bigr\}_{i=1}^{4} =
\bigl\{{{}^{1}p^{1+},{}^{4}\mbox{He}^{2+},{}^{14}\mbox{N}^{7+},{}^{56}\mbox{Fe}^{26+}}\bigr\}$
is considered. The nuclei correspond to the four groups of
elements which are the main components of cosmic rays with studied
energies \cite{PPNP-63-293-2009}. It should be noted that the free
parameter values in (\ref{eq:2.2})--(\ref{eq:2.5}) for $A+A$ have
been obtained for heavy ion collisions and usually for the most central bin. Consequently, the estimations derived within the present work are for most central collisions. In any case the
applicability of just the aforementioned analytic relations for light
nuclei requires the additional justification and careful
verification. Therefore the future estimations for light nucleus-nucleus collisions can be considered as preliminary with taking into account this feature.

Table \ref{tab:1} shows the $\sqrt{\smash[b]{s_{NN}}}$ values for the
laboratory energies of the incoming nuclei from the set
$\mathcal{G}_{Y}$ corresponding to some fixed ("nominal") values of
proton energy $E_{p}$. The UHECR with highest energies under
consideration allow the study of the final-state matter created in
PeV domain for $\sqrt{\smash[b]{s_{NN}}}$ which is far above any further
accelerator facilities projected now. As stressed above, the present analysis supposes the absence of the noticeable contributions of a new physics up to the $\sqrt{\smash[b]{s_{NN}}} \sim 1$ PeV.

\begin{table*}
\caption{\label{tab:1}Center-of-mass energies (TeV) for collisions with various
incoming nuclei.}
\begin{center}
\begin{tabular}{cccccccccc} \hline
\multicolumn{1}{c}{ Particle} & \multicolumn{9}{c}{$E_{p}$, TeV} \rule{0pt}{10pt}\\
\cline{2-10}
 & $10^{5}$ & $5 \times 10^{5}$ & $10^{6}$ & $5 \times 10^{6}$ & $10^{7}$ & $5 \times 10^{7}$ & $10^{8}$ & $5 \times 10^{8}$ & $10^{9}$ \rule{0pt}{10pt}\\
\hline
${}^{1}p^{1+}$           & 13.70 & 30.63 & 43.32 & 96.86 & 137.0 & 306.3 & 433.2 & 968.6 & 1370 \rule{0pt}{10pt}\\
${}^{4}\mbox{He}^{2+}$   & 9.686 & 21.66 & 30.63 & 68.49 & 96.86 & 216.6 & 306.3 & 684.9 & 968.6 \rule{0pt}{10pt}\\
${}^{14}\mbox{N}^{7+}$   & 9.686 & 21.66 & 30.63 & 68.49 & 96.86 & 216.6 & 306.3 & 684.9 & 968.6 \rule{0pt}{10pt}\\
${}^{56}\mbox{Fe}^{26+}$ & 9.334 & 20.87 & 29.52 & 66.00 & 93.34 & 208.7 & 295.2 & 660.0 & 933.4 \rule{0pt}{10pt}\\
\hline
\end{tabular}
\end{center}
\end{table*}

The estimations for global characteristics described above are
presented in Table \ref{tab:2} for three "nominal" values of $E_{p}$. The quantities $N_{\scriptsize{\mbox{ch}}}$,
$\rho^{E}_{\scriptsize{\mbox{Bj}}}$, $T$ and
$\tau_{\scriptsize{\mbox{kin}}}$ depend on $\langle
N_{\scriptsize{\mbox{part}}}\rangle$. The energy dependence of the
last parameter provides additional uncertainty for the global
characteristics under consideration especially in UHECR energy
domain and for light nuclei because of (very) limited data. In
order to avoid this source of dispersion the appropriate scale
factors are added for some global characteristics in Table
\ref{tab:2}. For the laboratory frame realized for UHECR passage
through the atmosphere the $\rho^{\eta}$ does not depend on $A$ and
$Z$ of the target nucleus and is the same for any asymmetric
$p+\mbox{A}$ collisions for given $E_{p}$. Therefore the values
for $p+\mbox{He}$ interactions are only shown on the second line
for $\rho^{\eta}$. The collisions of all considered types,
$p+p$, $p+A$, and $A+A$, are characterized by the large values of $\rho^{\eta}$
and scaled $N_{\scriptsize{\mbox{ch}}}$ for all nucleus from the set $\mathcal{G}_{Y}$.
For instance, the multiplicity density (\ref{eq:2.2}) in $p+p$ interactions at $E_{p}=10^{19}$ eV is already equal to the value of this parameter in most central $\mbox{Pb}+\mbox{Pb}$ collisions at $\sqrt{\smash[b]{s_{NN}}}=5.02$ TeV \cite{PRL-116-222302-2016}.
The values of $\rho^{E}_{\scriptsize{T}}$ and scaled parameter (\ref{eq:2.5})
at $\tau=1$ fm/$c$ are
extremely high for the medium created
in the final state of collisions of any nuclei from the $\mathcal{G}_{Y}$
already at $E_{p}=10^{17}$ eV. The Bjorken energy density with taking into account the scale factor is well above the estimation for the critical value of energy density $\rho^{E}_{c}=(0.34 \pm 0.16)$ GeV/fm$^{3}$ \cite{IJMPE-24-1530007-2015} for transition from the hadronic phase to the quark-gluon one for any nuclei and energies under study (Table \ref{tab:2}). The values of $\left.\xi^{-1}\rho^{E}_{\scriptsize{\mbox{Bj}}}\right|_{\tau=1}$ in small system collisions exceed significantly the value of the parameter in $\mbox{Pb}+\mbox{Pb}$ collisions at $\sqrt{\smash[b]{s_{NN}}}=2.76$ TeV which is $\approx 0.07$ GeV/fm$^{3}$ \cite{PRL-109-152303-2012}. Thus, one can expect the values for energy densities $\rho^{E}_{\scriptsize{T}}$ and $\rho^{E}_{\scriptsize{\mbox{Bj}}}$ correspond to the creation of the quark-gluon deconfined phase state in the collisions of the particles from the set $\mathcal{G}_{Y}$ at the energies under consideration. The matter created in the symmetric collisions of UHECR particles $\bigl\{\mathcal{G}_{Y}^{i}\bigr\}_{i=2}^{4}$ is characterized by the temperature larger significantly than the critical one $T_{c}=(156.5 \pm 1.5)$ MeV \cite{NPA-982-211-2019}. One can note the temperature can achieve the value which is of about 1 GeV in light nucleus-nucleus collisions at the highest center-of-mass energies corresponding to the $E_{p}= 10^{21}$ eV. In this case one can suggest that the non-perturbative effects in medium will be weaker than those for the matter studied in the collider experiments. Therefore the UHECR collisions for the $E_{p}$ range under study
create the quark-gluon matter which reaches the thermodynamic equilibrium and
lives the sufficiently long time. The last statement is confirmed by the
values of scaled $\tau_{\scriptsize{\mbox{kin}}}$ (Table \ref{tab:2}).
Furthermore the creation of the quark-gluon matter is already expected in
$\mbox{He}+\mbox{He}$ collisions at $\sqrt{\smash[b]{s_{NN}}}$ corresponding to the
low-boundary $E_{p}=10^{17}$ eV, i.e. the LHC domain.

\begin{table*}
\caption{\label{tab:2}Estimations
for global parameters in various collisions at some UHECR
energies.}
\begin{center}
\begin{tabular}{cccccc} \hline
\multicolumn{1}{c}{~Parameter~} & \multicolumn{1}{c}{~$E_{p}$, TeV~} &\multicolumn{1}{c} {~${}^{1}p^{1+}$~} &\multicolumn{1}{c}{~${}^{4}\mbox{He}^{2+}$~}&\multicolumn{1}{c}{~${}^{14}\mbox{N}^{7+}$~}&\multicolumn{1}{c}{~${}^{56}\mbox{Fe}^{26+}$~}\\
\hline
$\rho^{\eta}$ & $10^{5}$  & $6.6 \pm 0.7$  & $46.5 \pm 1.7$ & $46.5 \pm 1.7$ & $45.9 \pm 1.7$ \\
              &           & --             & $13.8 \pm 0.6$ & -- & --\\
              & $10^{7}$ & $11.1 \pm 1.4$ & $95 \pm 4$     & $95 \pm 4$     & $94 \pm 4$ \\
              &           & --             & $22.2 \pm 1.1$ & -- & --\\
              & $10^{9}$ & $18.8 \pm 2.6$ & $194 \pm 9$    & $194 \pm 9$    & $191 \pm 9$ \\
              &           & --             & $35.6 \pm 1.8$ & -- &
              -- \\
\hline
$\xi^{-1}N_{\scriptsize{\mbox{ch}}}$ & $10^{5}$ & $80.5 \pm 2.2$ & $141 \pm 20$ & $141 \pm 20$   & $139 \pm 20$ \\
            & $10^{7}$ & $154.7 \pm 2.3$ & $331 \pm 20$ & $331 \pm 20$ & $327 \pm 20$ \\
            & $10^{9}$ & $291.6 \pm 2.4$ & $768 \pm 21$ & $768 \pm 21$ & $758 \pm 21$ \\
\hline
$\rho^{E}_{\scriptsize{T}}$, & $10^{5}$ & -- & $18 \pm 7$ & $18 \pm 7$ & $18 \pm 7$ \\
    GeV     & $10^{7}$ & -- & $45 \pm 19$  & $45 \pm 19$  & $45 \pm 19$ \\
            & $10^{9}$ & -- & $110 \pm 50$ & $110 \pm 50$ & $110 \pm 50$ \\
\hline
$\left.\xi^{-1}\rho^{E}_{\scriptsize{\mbox{Bj}}}\right|_{\tau=1}$, & $10^{5}$ & --   & $1.5 \pm 0.6$  & $0.63 \pm 0.26$   & $0.24 \pm 0.10$ \\
  GeV/fm$^{3}$            & $10^{7}$ & --  & $3.7 \pm 1.6$ & $1.6 \pm 0.7$   & $0.62 \pm 0.27$ \\
                          & $10^{9}$ & --  & $9 \pm 4$     & $4.0 \pm 1.9$   & $1.6 \pm 0.7$ \\
\hline
$\left.\xi^{-1/4}T\right|_{\tau=1}$, & $10^{5}$  & -- & $0.55 \pm 0.06$  & $0.45 \pm 0.05$   & $0.35 \pm 0.04$ \\
    GeV                            & $10^{7}$ & -- & $0.70 \pm 0.08$ & $0.57 \pm 0.06$ & $0.45 \pm 0.05$ \\
                                   & $10^{9}$ & -- & $0.88 \pm 0.10$ & $0.71 \pm 0.08$ & $0.56 \pm 0.07$ \\
\hline
$\xi^{-1/3}\tau_{\scriptsize{\mbox{kin}}}$, & $10^{5}$ & -- & $13.55 \pm 0.17$ & $13.55 \pm 0.17$ & $13.40 \pm 0.17$ \\
   fm/$c$     & $10^{7}$ & -- & $27.7 \pm 0.4$ & $27.7 \pm 0.4$ & $27.4 \pm 0.4$ \\
              & $10^{9}$ & -- & $56.5 \pm 0.9$ & $56.5 \pm 0.9$ & $55.8 \pm 0.9$ \\
\hline
\end{tabular}
\end{center}
\end{table*}

The space-time extents of the emission region are evaluated for
$p+p$ collisions with the help of the results obtained for secondary
pions in \cite{Okorokov-AHEP-2016-5972709-2016}. The experimental
results for BE correlations for other nuclei from the set
$\mathcal{G}_{Y}$ are absent. Thus the estimations for BE
parameters in symmetric nuclear collisions for the subset
$\bigl\{\mathcal{G}_{Y}^{i}\bigr\}_{i=2}^{4}$ are calculated based
on the 3D analysis of the available experimental data for the
$\sqrt{\smash[b]{s_{NN}}}$ dependence of the scaled BE radii of the
charged-pion emission region created in various nuclear
interactions \cite{Okorokov-AHEP-2015-790646-2015}. Table
\ref{tab:3} summarizes the estimations for the space-time extents of
the source for secondary particle (pions) at some UHECR energies.
For given parameter and $E_{p}$ value the first-column values are
based on the results of approximations of the energy dependence of
femtoscopic radii by the general expression (\ref{eq:2.7}), while the
second columns show the estimations deduced with the help of the
results for the specific case of the fit function (\ref{eq:2.7}) at
$\forall\,i=s, o, l: a_{3}^{i}=1.0$ (fixed). The $p+p$ interactions produce the quasi-spherical source with equal radii within large errors, while the cylindrical shape of the emission region is clearly seen for nucleus-nucleus collisions with $R_{l} > R_{s}=R_{o}$ within uncertainties. Furthermore the excess of $R_{l}$ over the transverse-plane radii increases with collision energy. As seen in Table \ref{tab:3}, the final-state matter in collisions of UHECR particles with atmosphere occupies a noticeable volume at freeze-out even for lightest system interactions ($p+p$, $\mbox{He}+\mbox{He}$) at low boundary energy under consideration. The source radii in the symmetric ${}^{14}\mbox{N}^{7+}$ collisions are comparable in order of magnitude with the space-time extents of the emission region in $\mbox{Cu}+\mbox{Cu}$ collisions at RHIC energies $\sqrt{\smash[b]{s_{NN}}}=$62.4--200 GeV \cite{Okorokov-AHEP-2015-790646-2015}, especially for longitudinal axis. The similar relations are valid for the radii in $\mbox{Fe}+\mbox{Fe}$ collisions for ultra-high energy cosmic rays and heavy-ion ($\mbox{Au}+\mbox{Au}$) collisions at RHIC energies $\sqrt{\smash[b]{s_{NN}}}=$62.4--200 GeV \cite{Okorokov-AHEP-2015-790646-2015}. Thus, the growth of space
extents of the emission region with collision energy expected in
the case of the general view of (\ref{eq:2.7}) provides the radius
values for symmetric ${}^{56}\mbox{Fe}^{26+}$ collisions
similar to those in heavy-ion interactions at $\sqrt{\smash[b]{s_{NN}}}
\simeq 0.1$ TeV already at low boundary $E_{p}=10^{17}$ eV of the
energy domain studied. Therefore the estimations for the emission
region geometry shown in Table \ref{tab:3} especially for the general case of
(\ref{eq:2.7}) at energies close to the GZK limit $E_{p} \gtrsim
10^{19}$ eV prove the collisions of some UHECR light nuclei with
air as the source of bulk of the strongly interacting medium as
large as in the modern collider experiments with heavy-ion beams
at $\sqrt{\smash[b]{s_{NN}}} \simeq 0.1$ TeV.

\begin{table*}
\caption{\label{tab:3}Estimations
for space-time characteristic of pion source at some UHECR
energies.}
\begin{center}
\begin{tabular}{cccccccccc} \hline
\multicolumn{1}{c}{Particle} & \multicolumn{1}{c}{~$E_{p}$, TeV~} &\multicolumn{8}{c}{BE parameter} \rule{0pt}{10pt}\\
\cline{2-10}
 & & \multicolumn{2}{c}{$R_{s}$, fm} & \multicolumn{2}{c}{$R_{o}$, fm} & \multicolumn{2}{c}{$R_{l}$, fm} & \multicolumn{2}{c}{$V$, fm$^{3}$} \\
\hline
               & $10^{5}$ & $1.83 \pm 0.27$ & $1.7 \pm 0.7$ & $1.4 \pm 0.6$ & $1.4 \pm 0.7$ & $1.85 \pm 0.07$ & $1.82 \pm 0.17$ & $100 \pm 30$ & $80 \pm 60$ \rule{0pt}{10pt}\\
 ${}^{1}p^{1+}$& $10^{7}$ & $2.4 \pm 0.4$ & $2.1 \pm 0.7$ & $1.7 \pm 0.8$ & $1.7 \pm 0.9$ & $2.18 \pm 0.12$ & $1.98 \pm 0.20$ & $190 \pm 60$ & $130 \pm 90$ \rule{0pt}{10pt}\\
               & $10^{9}$ & $2.9 \pm 0.4$ & $2.5 \pm 0.9$ & $1.9 \pm 0.9$ & $1.9 \pm 1.1$ & $2.63 \pm 0.19$ & $2.14 \pm 0.24$ & $360 \pm 110$ & $210 \pm 140$ \rule{0pt}{10pt}\\
\hline
               & $10^{5}$ & $2.0 \pm 0.5$ & $1.60 \pm 0.07$ & $1.7 \pm 0.5$ & $1.72 \pm 0.07$ & $2.2 \pm 0.4$ & $2.25 \pm 0.09$ & $130 \pm 70$ & $91 \pm 9$ \rule{0pt}{10pt}\\
 ${}^{4}\mbox{He}^{2+}$& $10^{7}$ & $2.6 \pm 1.0$ & $1.71 \pm 0.07$ & $1.7 \pm 0.5$ & $1.78 \pm 0.08$ & $2.3 \pm 0.4$ & $2.50 \pm 0.11$ & $250 \pm 200$ & $115 \pm 11$ \rule{0pt}{10pt}\\
               & $10^{9}$ & $3.6 \pm 1.9$ & $1.81 \pm 0.08$ & $1.8 \pm 0.6$ & $1.84 \pm 0.08$ & $2.4 \pm 0.4$ & $2.75 \pm 0.12$ & $500 \pm 500$ & $142 \pm 14$ \rule{0pt}{10pt}\\
\hline
               & $10^{5}$ & $3.0 \pm 0.8$ & $2.43 \pm 0.10$ & $2.6 \pm 0.8$ & $2.62 \pm 0.11$ & $3.3 \pm 0.5$ & $3.42 \pm 0.14$ & $460 \pm 250$ & $320 \pm 30$ \rule{0pt}{10pt}\\
 ${}^{14}\mbox{N}^{7+}$& $10^{7}$ & $4.0 \pm 1.6$ & $2.60 \pm 0.11$ & $2.7 \pm 0.8$ & $2.70 \pm 0.12$ & $3.5 \pm 0.6$ & $3.80 \pm 0.16$ & $900 \pm 700$ & $400 \pm 40$ \rule{0pt}{10pt}\\
               & $10^{9}$ & $5.6 \pm 2.9$ & $2.75 \pm 0.12$ & $2.7 \pm 0.8$ & $2.79 \pm 0.13$ & $3.7 \pm 0.6$ & $4.17 \pm 0.18$ & $1800 \pm 1900$ & $500 \pm 50$ \rule{0pt}{10pt}\\
\hline
               & $10^{5}$ & $4.7 \pm 1.2$ & $3.86 \pm 0.16$ & $4.1 \pm 1.3$ & $4.16 \pm 0.18$ & $5.2 \pm 0.8$ & $5.42 \pm 0.23$ & $1800 \pm 1000$ & $1270 \pm 120$ \rule{0pt}{10pt}\\
 ${}^{56}\mbox{Fe}^{26+}$& $10^{7}$ & $6.3 \pm 2.5$ & $4.11 \pm 0.18$ & $4.2 \pm 1.3$ & $4.29 \pm 0.19$ & $5.5 \pm 0.9$ & $6.01 \pm 0.25$ & $3500 \pm 2800$ & $1600 \pm 150$ \rule{0pt}{10pt}\\
               & $10^{9}$ & $9 \pm 5$ & $4.36 \pm 0.19$ & $4.3 \pm 1.3$ & $4.42 \pm 0.20$ & $5.8 \pm 0.9$ & $6.62 \pm 0.28$ & $7000 \pm 7000$ & $1900 \pm 190$ \rule{0pt}{10pt}\\
\hline
\end{tabular}
\end{center}
\end{table*}

\begin{figure*}
\includegraphics[width=16.0cm,height=16.0cm]{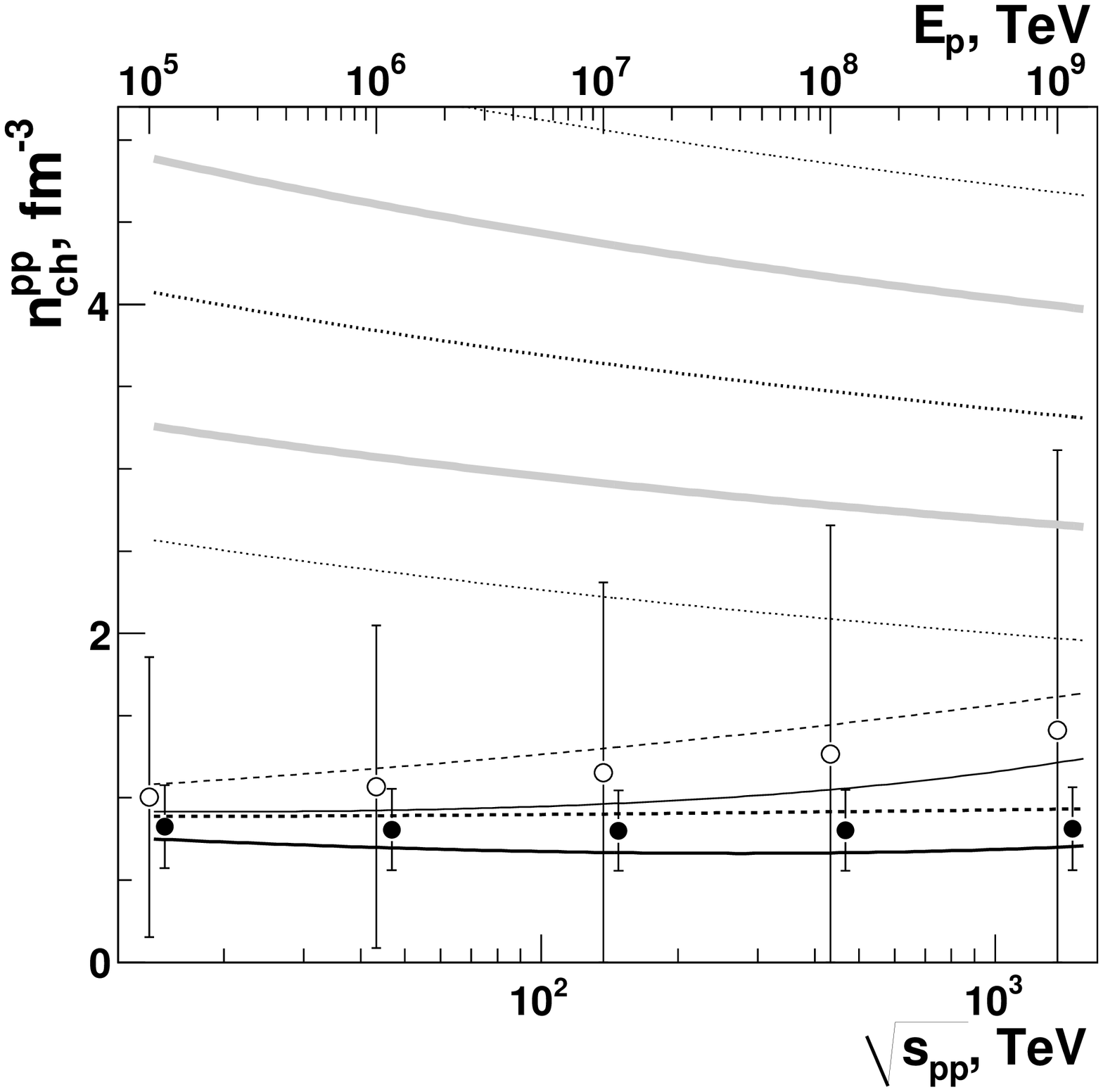}
\vspace*{8pt} \caption{
Energy
dependence of estimations for charged particle density and for
critical one in $p+p$ collisions calculated within the approach from \cite{Okorokov-AHEP-2016-5972709-2016}.
Points are calculated
with help of the power function (\ref{eq:2.3}) and the
estimations for $V$ based on the fits of BEC radii by (\ref{eq:2.7}) in the general
case ({\large$\bullet$}) and by
$R_{i} \propto \ln\varepsilon$, $i=s, l$ ({\large$\circ$}), while uncertainties for points are propagated
from statistical errors of fits used.
Solid lines correspond to the hybrid
approximation of $N_{\scriptsize{\mbox{ch}}}$ \cite{PRD-93-054046-2016} and
dashed lines are for 3NLO perturbative QCD equation
\cite{PR-349-301-2001}, while thick lines show results with
$V$ calculated with the fits of BEC radii by (\ref{eq:2.7}) in the general
case and thin lines -- with the fits by specific case
$R_{i} \propto \ln\varepsilon$, $i=s, l$. Critical
charged particle density is shown by dotted line with its
statistical uncertainty levels represented by thin dotted lines.
The heavy grey lines correspond to the systematic $\pm 1$ s.d. of
$n_{\mbox{\scriptsize{ch}}}$ calculated by varying of the fraction of the pions to be emitted from
a static Gaussian source on
$\pm 0.05$.} \label{fig:1}
\end{figure*}

\begin{figure*}
\includegraphics[width=16.0cm,height=16.0cm]{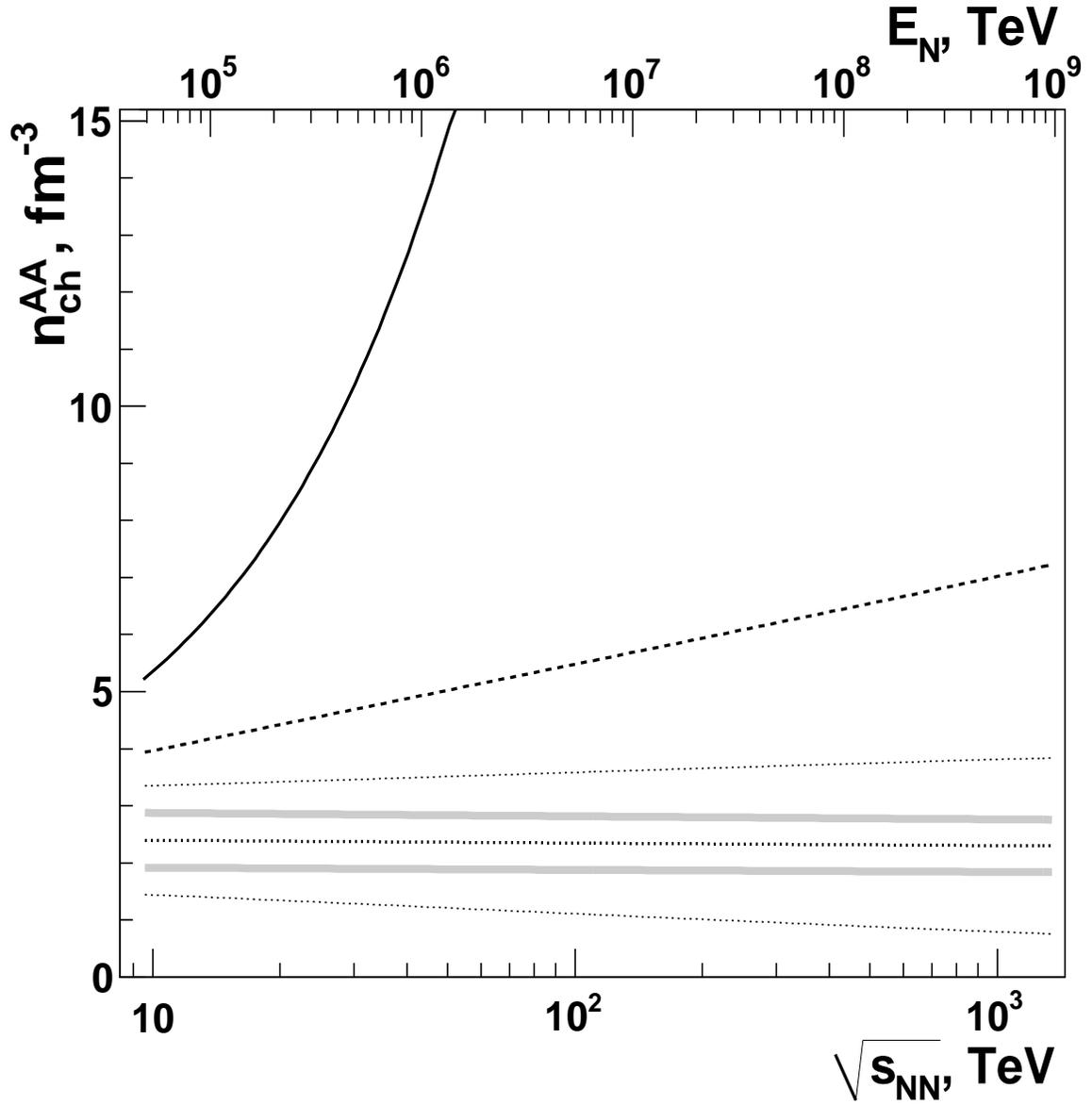}
\vspace*{8pt} \caption{Energy dependence of estimations for
charged particle density and for critical one in symmetric
($A+A$) heavy-ion collisions obtained within approach from \cite{Okorokov-AHEP-2016-5972709-2016}.
Solid line corresponds to the hybrid approximation of
$N_{\scriptsize{\mbox{ch}}}$ \cite{PRD-93-054046-2016} and dashed line is
for parametrization of total charged multiplicity from
\cite{PLB-726-610-2013}. Critical charged particle density
is shown by dotted line with its statistical uncertainty levels
represented by thin dotted lines. The heavy grey lines correspond
to the systematic $\pm 1$ s.d. of $n_{\mbox{\scriptsize{ch}}}$
calculated by varying of the fraction of the pions to be emitted from
a static Gaussian source on $\pm 0.05$.} \label{fig:2}
\end{figure*}

Global characteristics of the medium created in ultra-high energy
small system collisions studied here for $\bigl\{\mathcal{G}_{Y}^{i}\bigr\}_{i=1}^{4}$
and approach developed elsewhere
\cite{Okorokov-AHEP-2016-5972709-2016} allow the
investigation of lasing behavior for secondary pions in the
UHECR energy domain. The results of the calculations with help of the power function (\ref{eq:2.3}) at some fixed energies shown by symbols in Fig. \ref{fig:1} are added by the smooth dependencies of the charged particle density on $\sqrt{\smash[b]{s_{NN}}}$ derived with hybrid approximation of scaled $N_{\scriptsize{\mbox{ch}}}$ \cite{PRD-93-054046-2016} for the more complete picture. The hypothesis of the Bose--Einstein condensation corresponding to the lasing feature for pion production seems unfavorable in $p+p$ collisions even at highest $E_{p}=10^{21}$ eV (Fig. \ref{fig:1}). It cannot be possible to derive the estimations for light nucleus collisions due to the
unknown $\langle
N_{\scriptsize{\mbox{part}}}\rangle$. However, pion
laser regime is supported in Fig. \ref{fig:2} shown for heavy-ion collisions at ultra-high energies for completeness.

\section{Conclusions}\label{sect:4}

Summarizing the foregoing, one can draw the following conclusions.

The medium produced in the collisions of the UHECR particles with air
is characterized by high energy density at midrapidity and temperatures well above
the critical ones for the creation of the quark-gluon
matter already at $\sqrt{\smash[b]{s_{NN}}}$ corresponding to the $E_{p} = 10^{17}$ eV.
BE decoupling time is about 10 fm/$c$ on order of value even in helium nucleus
collisions. Therefore for the first time the quantitative analysis of the wide set of global and geometric characteristics strongly
indicates that the long-lived medium in quark-gluon phase can be produced in
light nuclei collisions at UHECR energies. The particle source created in small system collisions
at  $E_{p} \gtrsim 10^{19}$ eV is characterized by the large
space-time extents which support the hypothesis of the creation of
blobs of the quark-gluon matter under extreme conditions in UHECR interactions.

Future experimental and theoretical study for collisions of light nuclei at ultra-high energies
are important for the verification of the estimations obtained within the paper.

\section*{Acknowledgments}

This work was supported in part within the Program
for Improving the Competitive Ability of National
Research Nuclear University MEPhI (Contract
no. 02.a03.21.0005 of August 27, 2013).

\newpage

\newpage

\end{document}